\documentclass[11pt,a4paper,english,nofootinbib]{revtex4}
\usepackage{lmodern}

\usepackage[T1]{fontenc}
\usepackage[latin9]{inputenc}
\usepackage{color}
\usepackage{babel}
\usepackage{amsmath}
\usepackage{amssymb}
\usepackage{esint}
\usepackage[unicode=true,pdfusetitle,
 bookmarks=true,bookmarksnumbered=false,bookmarksopen=false,
 breaklinks=false,pdfborder={0 0 1},backref=false,colorlinks=false]
 {hyperref}

\makeatletter

\pdfpageheight\paperheight
\pdfpagewidth\paperwidth

\@ifundefined{textcolor}{}
{%
 \definecolor{BLACK}{gray}{0}
 \definecolor{WHITE}{gray}{1}
 \definecolor{RED}{rgb}{1,0,0}
 \definecolor{GREEN}{rgb}{0,1,0}
 \definecolor{BLUE}{rgb}{0,0,1}
 \definecolor{CYAN}{cmyk}{1,0,0,0}
 \definecolor{MAGENTA}{cmyk}{0,1,0,0}
 \definecolor{YELLOW}{cmyk}{0,0,1,0}
}

\usepackage{latexsym}\usepackage{bm}

\makeatother

\definecolor{green}{RGB}{0, 180, 0}
\definecolor{cyan}{RGB}{0, 180, 180}
\definecolor{yellow}{RGB}{211,211,0}

\makeatother

\begin{document}

\title{Entropy formula in Einstein-Maxwell-Dilaton theory and its validity for black strings}

\author{Mohammad Reza Setare}
\email{rezakord@ipm.ir}

\affiliation{{Department of Science, Campus of Bijar, University of Kurdistan, Bijar, Iran }}

\author{Hamed Adami}
\email{hamed.adami@yahoo.com}

\affiliation{{Research Institute for Astronomy and Astrophysics of Maragha (RIAAM),
P.O. Box 55134-441, Maragha, Iran }}

\begin{abstract}
We consider near horizon fall-off conditions of stationary black holes in Einstein-Maxwell-Dilaton theory and find conserved charge conjugate to symmetry generator that preserves near horizon fall-off conditions. Subsequently, we find supertranslation, superrotation and multiple-charge modes. We apply the obtained results on a typical static dilaton black hole and on a charged rotating black string, as examples. In this case, supertranslation double-zero-mode charge $\mathcal{T}_{(0,0)}$ is not equal to black hole entropy times Hawking temperature. This may be seen as a problem but it is not, because, in Einstein-Maxwell-Dilaton theory, we have a U(1) gauge freedom and we use an appropriate gauge fixing to fix that problem. We show that new entropy formula $4 \pi \hat{J}^{+}_{0} \hat{J}^{-}_{0}$, proposed in \cite{17}, is valid for black strings as well as black holes.
\end{abstract}

\maketitle

\section{Introduction}\label{S.I}
The Einstein-Maxwell-dilaton (EMD) theory originating from a low energy
limit of string theory, allows for black holes that
have mass, rotation, charge and scalar hair \cite{25}. The uniqueness of static, asymptotically flat spacetimes with non-degenerate
black holes of Einstein-Maxwell-dilaton theory has been investigated in \cite{26}. The dilaton
field can change the asymptotic behavior of the solutions to be neither asymptotically flat nor (A)dS. Rotating solutions of EMD gravity with Liouville-type potential in four and (n + 1)-dimensions when horizon is flat
have been studied respectively in \cite{27,28}. These solutions \cite{27,28}  describe charged rotating dilaton
black strings/branes.\\
In this paper we would like to provide the first non-trivial evidence for universality of the entropy formula $4\pi J_{0}^{+}J_{0}^{-}$ in Einstein-Maxwell-dilaton gravity in 4-dimensions. Recently the above entropy formula emerged in the near horizon description of non-extermal Kerr black holes in 4-dimensions \cite{17}. In previous paper \cite{7} we have show that this entropy formula give us the correct results for Kerr-Newman (A)dS black holes. In order to investigate universality of the above entropy formula, here we study the near horizon fall-off conditions of stationary black holes in Einstein-Maxwell-Dilaton gravity. We show that the above new entropy formula not only is valid for black hole solutions of this theory but also work correctly for black string solutions. For this propose we use the covariant phase space method for obtaining conserved charges in EMD theory. Then we study the near horizon behavior of a stationary black hole in EMD theory. We find conserved charge conjugate to symmetry generator that preserves near horizon fall-off conditions. After that we obtain supertranslation, superrotation and multiple-charge modes. 
\section{Conserved charges in Einstein-Maxwell-Dilaton Theory}\label{S.II}
Let us consider Einstein-Maxwell-Dilaton (EMD) theory. The Lagrangian describing EMD theory is a functional of metric $g_{\mu \nu}$, $U(1)$ gauge field $A_{\mu}$ and a real scalar field $\varphi$
 \begin{equation}\label{1}
  L=\sqrt{-g} \left( \mathcal{R}-2 \Lambda- \frac{1}{2} \partial_{\mu} \varphi \partial^{\mu} \varphi - \frac{1}{4} \mathcal{N}(\varphi) F_{\mu \nu} F^{\mu \nu} - \mathcal{V}(\varphi)\right),
\end{equation}
where $\mathcal{R}$, $F_{\mu \nu} = \partial _{\mu} A_{\nu} - \partial _{\nu} A_{\mu} $ and $\Lambda$ are respectively the Ricci scalar, electromagnetic field strength and the cosmological constant.\\
Now, we briefly review the approach of the covariant phase space method for obtaining conserved charges in EMD theory. To do this, we follow references \cite{1,2,3,4,5}. First order variation of the Lagrangian \eqref{1} is
\begin{equation}\label{2}
  \delta L [\Phi]=E_{\Phi}[\Phi] \delta \Phi + \partial_{\mu} \Theta ^{\mu}[\Phi , \delta \Phi],
\end{equation}
where $\Phi = \{ g_{\mu \nu} , A_{\mu}, \varphi \}$ denotes collection of dynamical fields, $E_{\Phi}$ have dual indices with $\Phi$ and sum on $\Phi$ is explicitly assumed. In the equation \eqref{2},
\begin{equation}\label{3}
  \Theta^{\mu}[\Phi, \delta \Phi]= \sqrt{-g} \left\{2 \nabla ^{[\alpha} \left( g^{\mu] \beta} \delta g_{\alpha \beta} \right)- \partial^{\mu} \varphi \delta \varphi -\mathcal{N} F^{\mu \nu} \delta A_{\nu} \right\},
\end{equation}
is the surface term and we refer to it as symplectic potential. Also, $E_{\Phi}=0$ give us the field equations
\begin{equation}\label{4}
   G^{\mu \nu} + \Lambda g^{\mu \nu} = T_{(A)}^{\mu \nu}+T_{(\varphi)}^{\mu \nu} ,
\end{equation}
\begin{equation}\label{5}
   \nabla _{\nu} \left( \mathcal{N} F^{\nu \mu} \right)=0,
\end{equation}
\begin{equation}\label{6}
  \Box \varphi - \mathcal{V}^{\prime}-\frac{1}{4} \mathcal{N}^{\prime} F_{\alpha \beta} F^{\alpha \beta}=0,
\end{equation}
where $G^{\mu \nu}$ is the Einstein tensor and
\begin{equation}\label{7}
  T_{(A)}^{\mu \nu} = \frac{1}{2} \mathcal{N} \left( F^{\mu \alpha} F^{\nu}_{\hspace{1.7 mm} \alpha} - \frac{1}{4} g^{\mu \nu} F^{\alpha \beta} F_{\alpha \beta} \right),
\end{equation}
\begin{equation}\label{8}
  T_{(\varphi)}^{\mu \nu} = \frac{1}{2} \partial^{\mu} \varphi \partial^{\nu} \varphi - \frac{1}{2} g^{\mu \nu} \left(\frac{1}{2} \partial_{\alpha} \varphi \partial^{\alpha} \varphi + \mathcal{V}\right),
\end{equation}
are contributions of electromagnetic field and scalar field in energy-momentum tensor, respectively. Here, the prime symbol denotes differentiation with respect to $\varphi$.\\
Now consider two arbitrary variations $\delta_{1}$ and $\delta_{2}$. Suppose these two variations do not commute $\delta_{1} \delta_{2} \neq \delta_{2} \delta_{1}$. By varying Eq.\eqref{2}, we find second order variation of the Lagrangian
\begin{equation}\label{9}
  \delta_{1} \delta_{2} L [\Phi]=\delta_{1} E_{\Phi}[\Phi] \delta_{2} \Phi +  E_{\Phi}[\Phi] \delta_{1} \delta_{2} \Phi + \partial_{\mu} \delta_{1} \Theta ^{\mu}[\Phi , \delta _{2} \Phi].
\end{equation}
Similarly, one can write
\begin{equation}\label{10}
\delta_{2} \delta_{1} L [\Phi]=\delta_{2} E_{\Phi}[\Phi] \delta_{1} \Phi +  E_{\Phi}[\Phi] \delta_{2} \delta_{1} \Phi + \partial_{\mu} \delta_{2} \Theta ^{\mu}[\Phi , \delta _{1} \Phi].
\end{equation}
By subtracting Eq.\eqref{10} from Eq.\eqref{9}, we have
\begin{equation}\label{11}
  \delta_{[1,2]} L [\Phi]=\delta_{1} E_{\Phi}[\Phi] \delta_{2} \Phi - \delta_{2} E_{\Phi}[\Phi] \delta_{1} \Phi +  E_{\Phi}[\Phi] \delta_{[1,2]} \Phi + \partial_{\mu} \left( \delta_{1} \Theta ^{\mu}[\Phi , \delta _{2} \Phi] - \delta _{2} \Theta ^{\mu}[\Phi , \delta _{1} \Phi] \right),
\end{equation}
where $\delta_{[1,2]}=\delta_{1} \delta_{2} - \delta_{2} \delta_{1}$ is commutator of two variations $\delta_{1}$ and $\delta_{2}$. By using Eq.\eqref{2}, and replacing $\delta \rightarrow \delta_{[1,2]} $, we can write Eq.\eqref{11} as
\begin{equation}\label{12}
  \partial_{\mu} \omega_{\text{LW}}^{\mu}[\Phi; \delta_{1} \Phi , \delta_{2} \Phi]=- \frac{1}{16 \pi} \left( \delta_{1} E_{\Phi}[\Phi] \delta_{2} \Phi - \delta_{2} E_{\Phi}[\Phi] \delta_{1} \Phi \right),
\end{equation}
where
\begin{equation}\label{13}
   \omega_{\text{LW}}^{\mu}[\Phi; \delta_{1} \Phi , \delta_{2} \Phi]= \frac{1}{16 \pi} \left( \delta_{1} \Theta ^{\mu}[\Phi , \delta _{2} \Phi] - \delta _{2} \Theta ^{\mu}[\Phi , \delta _{1} \Phi] - \Theta ^{\mu}[\Phi , \delta _{[1,2]} \Phi] \right),
\end{equation}
is the Lee-Wald symplectic current. Since the symplectic potential is linear in $\delta \Phi$ then the terms containing $\delta _{1} \Phi$, $\delta _{2} \Phi$ and $\delta _{[1,2]} \Phi$ eliminate each other and $\omega_{\text{LW}}^{\mu}$ is a skew-symmetric bilinear in $\delta _{1} \Phi$ and $\delta _{2} \Phi$. The Lee-Wald symplectic current is conserved when equations of motion and linearized equations of motion are satisfied. In other words, if $\Phi$ is a solution of $E_{\Phi}=0$ and $\delta _{1} \Phi$ and $\delta _{2} \Phi$ are solutions of $\delta E_{\Phi}=0$, then the Lee-Wald symplectic current is conserved
\begin{equation}\label{14}
  \partial_{\mu} \omega_{\text{LW}}^{\mu}[\Phi; \delta_{1} \Phi , \delta_{2} \Phi] \simeq 0.
\end{equation}
The sign $\simeq$ indicates that the equality is held on-shell. We can define symplectic 2-form on solution space through the Lee-Wald symplectic current
\begin{equation}\label{15}
  \Omega_{\text{LW}} [\Phi; \delta_{1} \Phi , \delta_{2} \Phi]= \int_{\mathcal{C}} \omega_{\text{LW}}^{\mu}[\Phi; \delta_{1} \Phi , \delta_{2} \Phi] d^{3} x_{\mu},
\end{equation}
where $\mathcal{C}$ is a codimension-1 spacelike surface. Solution phase space can be constructed by factoring out the degeneracy subspace of configuration space (see Ref.\cite{1} for detailed discussion). Hence $\Omega_{\text{LW}}$ will be a symplectic form on solution phase space and it is closed, skew-symmetric and non-degenerate.\\
Suppose $\xi^{\mu}(x)$ and $\lambda(x)$ to be generators of diffeomorphism and $U(1)$ gauge transformation. We can introduce a combined transformation so that $\chi= (\xi , \lambda)$ is the generator of such transformations \cite{6,7}. The change in metric, $U(1)$ gauge field and scalar field induced by an infinitesimal transformation generated by $\chi$ are given by
\begin{equation}\label{16}
  \delta _{\chi} g_{\mu \nu}= \mathcal{L}_{\xi} g_{\mu \nu},
\end{equation}
\begin{equation}\label{17}
  \delta _{\chi} A_{\mu}= \mathcal{L}_{\xi} A_{\mu} + \partial _{\mu} \lambda,
\end{equation}
\begin{equation}\label{18}
  \delta _{\chi} \varphi= \mathcal{L}_{\xi} \varphi,
\end{equation}
respectively. Here, $\mathcal{L}_{\xi}$ denotes the Lie derivative along the vector field $\xi$. Also, the change in Lagrangian \eqref{1} induced by an infinitesimal transformation generated by $\chi$ is
\begin{equation}\label{19}
  \delta _{\chi} L[\Phi]= \mathcal{L}_{\xi} L[\Phi]= \partial_{\mu}\left( \xi^{\mu} L[\Phi] \right).
\end{equation}
Since change in dynamical fields are linear in generator $\chi$ and change in the Lagrangian is a total derivative then $\chi$ generates a local symmetry on solution phase space \cite{1}. The generators of such local symmetry on solution phase space are conserved charges. The charge perturbation conjugate to $\chi$ is defined as
\begin{equation}\label{20}
  \delta Q_{\chi}= \Omega_{\text{LW}} [\Phi; \delta \Phi , \delta_{\chi} \Phi].
\end{equation}
The algebra among conserved charges is
\begin{equation}\label{21}
  \{ Q_{\chi_{1}}, Q_{\chi_{2}} \} = Q_{[\chi_{1},\chi_{2}]}+\tilde{\mathcal{C}}(\chi_{1},\chi_{2}),
\end{equation}
where $\tilde{\mathcal{C}}(\chi_{1},\chi_{2})$ is central extension term and the Dirac bracket is defined as
\begin{equation}\label{22}
  \{ Q_{\chi_{1}}, Q_{\chi_{2}} \} = \delta_{\chi_{2}}Q_{\chi_{1}} .
\end{equation}
Now, we want to find explicit form of conserved charges in the EMD theory. To this end, we assume that the variation in Eq.\eqref{1} is induced by an infinitesimal transformation generated by $\chi$
\begin{equation}\label{23}
  \delta _{\chi} L [\Phi] \simeq \partial_{\mu} \Theta ^{\mu}[\Phi , \delta _{\chi} \Phi],
\end{equation}
then we can define an on-shell Noether current
\begin{equation}\label{24}
    J_{\text{N}}^{\mu}[\Phi ; \chi] \simeq \Theta ^{\mu}[\Phi , \delta _{\chi} \Phi]- \xi^{\mu} L[\Phi],
\end{equation}
which is conserved on-shell, i.e. $\partial _{\mu}J_{\text{N}}^{\mu} \simeq 0 $. Thus there exists a second rank tensor density of weight $+1$
\begin{equation}\label{25}
  K_{\text{N}}^{\mu \nu}[\Phi ; \chi] \simeq -  \sqrt{-g} \left\{ 2 \nabla ^{[\mu} \xi^{\nu]} +\mathcal{N} F^{\mu \nu} (\lambda + A_{\alpha} \xi^{\alpha}) \right\},
\end{equation}
so that $ J_{\text{N}}^{\mu} \simeq \partial _{\nu} K_{\text{N}}^{\mu \nu}$. We refer to $K_{\text{N}}^{\mu \nu}$ as Noether potential. To find explicit form of the symplectic current, first, we take an arbitrary variation from Eq.\eqref{24}
\begin{equation}\label{26}
  \partial _{\nu} \delta K_{\text{N}}^{\mu \nu}[\Phi ; \chi] \simeq \delta \Theta ^{\mu}[\Phi , \delta _{\chi} \Phi]- \delta \left( \xi^{\mu} L[\Phi] \right).
\end{equation}
To have generality we assume that $\chi$ depends on the dynamical fields. On the other hand, second variation of Eq.\eqref{1}, induced by an infinitesimal transformation generated by $\chi$, is
\begin{equation}\label{27}
  \delta _{\chi} \delta L [\Phi] \simeq \partial_{\mu} \delta _{\chi} \Theta ^{\mu}[\Phi , \delta \Phi].
\end{equation}
 Since the commutator of an arbitrary variation and a variation induced by an infinitesimal transformation generated by $\chi$ is $ \delta \delta _{\chi}-\delta _{\chi} \delta = \delta _{\delta \chi}$ then the equation \eqref{27} can be written as
\begin{equation}\label{28}
  \delta \delta_{\chi} L [\Phi] - \delta _{\delta \chi} L [\Phi] \simeq \partial_{\mu} \delta _{\chi} \Theta ^{\mu}[\Phi , \delta \Phi].
\end{equation}
By substituting Eq.\eqref{23} into Eq.\eqref{28}, we find the explicit form of the symplectic current as
\begin{equation}\label{29}
    \omega_{\text{LW}}^{\mu} [\Phi; \delta \Phi , \delta_{\chi} \Phi] \simeq  \partial_{\nu} \mathcal{Q}_{\text{LW}}^{\mu \nu} [\Phi, \delta \Phi ;\chi ],
\end{equation}
with
\begin{equation}\label{30}
    \mathcal{Q}_{\text{LW}}^{\mu \nu} [\Phi, \delta \Phi ;\chi ] =\frac{1}{16 \pi} \left\{ \delta K_{\text{N}}^{\mu \nu}[\Phi ; \chi] - \delta K_{\text{N}}^{\mu \nu}[\Phi ; \delta \chi] + 2 \xi ^{[\mu} \Theta ^{\nu]}[\Phi , \delta \Phi ] \right\}.
\end{equation}
In the EMD theory the explicit form of $\mathcal{Q}_{\text{LW}}^{\mu \nu}$ can be found by substituting equations \eqref{25} and \eqref{3} into the above equation
\begin{equation}\label{31}
\begin{split}
   \mathcal{Q}_{\text{LW}}^{\mu \nu} [\Phi, \delta \Phi ;\chi ] =  \frac{\sqrt{-g}}{8 \pi} \biggl\{ & h^{\lambda [ \mu} \nabla _{\lambda} \xi ^{\nu]} - \xi ^{\lambda} \nabla ^{[\mu} h^{\nu]}_{\lambda} - \frac{1}{2} h \nabla ^{[\mu} \xi ^{\nu]} + \xi ^{[\mu} \nabla _{\lambda} h^{\nu] \lambda} - \xi ^{[\mu} \nabla^{\nu]}h \\
     & - \frac{1}{2} \left[ \mathcal{N} \left( \delta F^{\mu \nu} + \frac{1}{2} h F^{\mu \nu} \right)+\mathcal{N}^{\prime} F^{\mu \nu} \delta \varphi \right] ( \lambda +A_{\alpha} \xi^{\alpha})\\
     & -\frac{1}{2} \mathcal{N} F^{\mu \nu} \xi^{\alpha} \delta A_{\alpha} - \mathcal{N} \xi^{[\mu} F^{\nu] \alpha} \delta A_{\alpha}- \xi^{[ \mu} \partial^{\nu ]} \varphi \delta \varphi \biggr\},
\end{split}
\end{equation}
where $h_{\mu \nu} = \delta g_{\mu \nu}$. We can use Eq.\eqref{29} and Stokes' theorem to write conserved charge perturbation \eqref{20} as
\begin{equation}\label{32}
  \delta Q_{\chi}= \oint_{\mathcal{D}} \mathcal{Q}_{\text{LW}}^{\mu \nu} [\Phi, \delta \Phi ;\chi ] d^{2}x_{\mu \nu} ,
\end{equation}
where $\mathcal{D}$ denotes boundary of $\mathcal{C}$ and it is a spacelike codimension-2 surface. Usually it is thought that the linearization is just valid at spatial infinity. To overcome this problem, we take an integration from Eq.\eqref{32} over one-parameter path on the solution phase space. To this end, suppose that $\Phi (\mathcal{W})$ is a collection of fields which solve the equations of motion of the EMD theory, where $\mathcal{W}$ is a free parameter in the solution phase space. Now, we replace $\mathcal{W}$ by $s\mathcal{W}$, where $0 \leq s \leq 1$ is just a parameter. By expanding $\Phi (s\mathcal{ W})$ in terms of $s$ we have $\Phi (s\mathcal{ W}) =  \Phi (0) + s \frac{\partial \Phi}{ \partial s} \big|_{s=0}  + \cdots$. By substituting $\Phi=\Phi (s\mathcal{ W})$ and $\delta \Phi =ds \frac{\partial \Phi}{ \partial s} \big|_{s=0}$ into Eq.\eqref{32}, we can define the conserved charge conjugate to $\chi$. Then we will have
\begin{equation}\label{33}
   Q_{\chi}= \int_{0}^{1} ds \oint_{\mathcal{D}} \mathcal{Q}_{\text{LW}}^{\mu \nu} [\Phi ;\chi | s ] d^{2}x_{\mu \nu} ,
\end{equation}
where integration over $s$ denotes integration over the one-parameter path on the solution phase space. In the equation \eqref{33}, $s=0$ is the value of the parameter corresponds to the background configuration. In this way, background contribution in the conserved charge is subtracted and then the conserved charge will be always finite. Therefore, this method is applicable to spacetimes with any backgrounds.
\section{Near horizon fall-off conditions and symmetries}\label{S.III}
Let us consider near horizon behavior of a stationary black hole in EMD theory. One can write near horizon metric in the Gaussian null coordinate system \cite{7,8,9,10}
\begin{equation}\label{34}
  ds^{2}= -2 \kappa \rho dv^{2}+2 dv d\rho + 2 \rho \theta_{A} dv dx^{A} +(\Omega_{AB}+ \rho \lambda_{AB}) dx^{A} dx^{B} + \mathcal{O}(\rho^{2}),
\end{equation}
where $v$ is the advanced time coordinate such that a null surface is defined by $g^{\alpha \beta} \partial_{\alpha}v \partial_{\beta}v =0$ and the vector tangent to this surface is given by $k^{\mu} =g^{\mu \nu} \partial_{\nu}v$ which defines a ray. Also, $\rho$ is the affine parameter of the
generator $k^{\mu}$. We assume that the horizon is located at $\rho=0$. Suppose $\kappa$, $\theta_{A}$, $\Omega_{AB}$ and $\lambda_{AB}$ are functions of $x^{A}$, where two coordinates $x^{A}$ are chosen constant along each ray. One can introduce following near horizon fall-off conditions for the $U(1)$ gauge field and scalar field
\begin{equation}\label{35}
  A_{v}=A_{v}^{(0)}+ A_{v}^{(1)} \rho + \mathcal{O}(\rho^{2}), \hspace{0.7 cm} A_{\rho}=0,\hspace{0.7 cm} A_{B}=A_{B}^{(0)}+ A_{B}^{(1)} \rho + \mathcal{O}(\rho^{2}),
\end{equation}
\begin{equation}\label{36}
  \varphi = \varphi ^{(0)}+\varphi^{(1)} \rho + \mathcal{O}(\rho^{2}),
\end{equation}
respectively, where we set $A_{\rho}=0$ as a gauge condition and $A_{v}^{(i)}$, $A_{B}^{(i)}$ and $\varphi ^{(i)}$ are functions of $x^{A}$.\\
By substituting fall-off conditions \eqref{34}, \eqref{35} and \eqref{36} into the field equations, we can find additional restrictions. The $(v,v) $ and  $(v,A) $ components of the equations of motion \eqref{4} at zeroth order restrict $\kappa$ and $A_{v}^{(0)}$ to be constants, i.e. independent of $x^{A}$. The other components of the equations of motion relate first order terms to zeroth order ones in metric, gauge field and scalar field expansions and we do not need them later.\\
In order to obtain the asymptotic symmetry generator $\chi$, we assumed that the leading terms does not depend on the dynamical fields. Under such an assumption the boundary conditions will be "state independent", which means that the form of the asymptotic symmetry generators are not considered to depend explicitly of the charges \cite{9}. In this way we can find components of symmetry generator $\chi$ as follows \cite{7,9}:
\begin{equation}\label{37}
\begin{split}
     & \xi^{v}= T, \hspace{1 cm} \xi^{\rho}= \frac{1}{2} \rho^{2} \Omega^{AB} \theta_{A} \partial _{B} T+ \mathcal{O}(\rho^{3}), \\
     & \xi^{A}= Y^{A} - \rho \Omega^{A B} \partial _{B} T+\frac{1}{2} \rho^{2} \Omega^{AC} \Omega^{BD} \lambda_{C D} \partial _{B} T+ \mathcal{O}(\rho^{3}),
\end{split}
\end{equation}
\begin{equation}\label{38}
  \lambda= \lambda^{(0)}+ \rho \Omega^{AB} \varphi_{A} \partial _{B} T - \frac{1}{2} \rho^{2} \left( \Omega^{AC} \Omega^{BD} \lambda_{C D} \varphi_{A} \partial _{B} T- \Omega^{AB}\psi_{A} \partial _{B} T\right) + \mathcal{O}(\rho^{3}),
\end{equation}
The symmetry generator $\chi$, with above components, preserves the given near horizon fall-off conditions. Here $T$, $Y^{A}$ and $\lambda^{(0)}$ are arbitrary functions of $x^{A}$. The change in dynamical fields under the action of symmetry generator $\chi$ can be read as
\begin{equation}\label{39}
  \begin{split}
       & \delta_{\chi} \theta_{A}= \mathcal{L}_{Y} \theta_{A}-2 \kappa \partial_{A} T, \hspace{1 cm}  \delta_{\chi} \Omega_{AB} =\mathcal{L}_{Y}\Omega_{AB} , \\
       & \delta_{\chi} \lambda_{AB} = \mathcal{L}_{Y} \lambda_{AB} + \theta_{A} \partial _{B} T+ \theta_{B} \partial _{A} T - 2 \bar{\nabla}_{A}\bar{\nabla}_{B} T,
  \end{split}
\end{equation}
\begin{equation}\label{40}
  \begin{split}
       & \delta_{\chi} A_{v}^{(1)}= \mathcal{L}_{Y}  A_{v}^{(1)}, \hspace{1 cm}  \delta_{\chi} A_{A}^{(0)} =\mathcal{L}_{Y} A_{A}^{(0)}+ A_{v}^{(0)} \partial_{A} T+ \partial_{A} \lambda^{(0)}, \\
       & \delta_{\chi} A_{A}^{(1)} = \mathcal{L}_{Y} A_{A}^{(1)} + A_{v}^{(1)} \partial_{A} T + \Omega^{BC} \left( \partial_{A} A_{B}^{(0)}-\partial_{B} A_{A}^{(0)}\right) \partial_{C}T,
  \end{split}
\end{equation}
\begin{equation}\label{41}
  \delta_{\chi} \varphi^{(0)}=\mathcal{L}_{Y} \varphi^{(0)},
\end{equation}
where $\mathcal{L}_{Y}$ denotes the Lie derivative along $Y^{A}$ and $\bar{\nabla}_{A}$ is the covariant derivative with respect to connection $\bar{\Gamma}^{A}_{BC}$ compatible with the metric of the horizon $\Omega_{AB}$. It is worth mentioning that because $\kappa$ and $A_{v}^{(0)}$ are not dynamical then they will remain unchanged under the action of the symmetry generator $\chi$, i.e. $\delta_{\chi}\kappa =0$ and $\delta_{\chi}A_{v}^{(0)}=0$.\\
The asymptotic Killing vectors \eqref{37} are functions of the dynamical fields. To take it into account we introduce a modified version of Lie brackets \cite{11}
\begin{equation}\label{42}
  \left[ \xi_{1} , \xi_{2} \right] =  \mathcal{L} _{\xi_{1}} \xi_{2} - \delta ^{(g)}_{\xi _{1}} \xi_{2} + \delta ^{(g)}_{\xi _{2}} \xi_{1},
\end{equation}
where $\xi_{1}= \xi(T_{1}, Y^{A}_{1})$ and $\xi_{2}= \xi(T_{2}, Y^{A}_{2})$ and $\delta ^{(g)}_{\xi _{1}} \xi_{2}$ denotes the change induced in $\xi_{2}$ due to the variation of metric $\delta _{_{\xi _{1}}} g_{\mu\nu} = \mathcal{L} _{\xi_{1}} g_{\mu\nu}$. Therefore one finds that
\begin{equation}\label{43}
  \left[ \xi_{1} , \xi_{2} \right] = \xi_{12},
\end{equation}
with $\xi_{12}= \xi(T_{12}, Y^{A}_{12})$, where
\begin{equation}\label{44}
  T_{12}= Y_{1}^{A} \partial_{A}T_{2} - Y_{2}^{A} \partial_{A}T_{1}, \hspace{1 cm} Y_{12}^{A}= Y_{1}^{B} \partial_{B}Y_{2}^{A}- Y_{2}^{B} \partial_{B}Y_{1}^{A}.
\end{equation}
Thus, the algebra of asymptotic Killing vectors is closed. In addition to $T$ and $Y^{A}$, the symmetry generator $\chi=\chi (T, Y^{A}, \lambda^{(0)})$ contains another degree of freedom, $\lambda^{(0)}$. Here, $\lambda^{(0)}$ is an arbitrary function of $x^{A}$ and generates $U(1)$ symmetry. Hence, we need to introduce two other commutators \cite{7}
\begin{equation}\label{45}
  [\chi(0,0,0,\lambda_{1}^{(0)}), \chi(0,0,0,\lambda_{2}^{(0)})]= 0,
\end{equation}
\begin{equation}\label{46}
  [\chi(0,0,0,\lambda_{1}^{(0)}), \chi(0,Y_{2}^{A},0)] =-[\chi(0,Y_{2}^{A},0),\chi(0,0,0,\lambda_{1}^{(0)})]=\chi(0,0,0, -\mathcal{L}_{Y_{2}} \lambda_{1}^{(0)}),
\end{equation}
in addition to Eq.\eqref{43}. The equation \eqref{45} comes from the fact that $U(1)$ is an Abelian group and we will justify Eq.\eqref{46} when we consider the algebra among conserved charges. In a nutshell, algebra among asymptotic symmetry generators can be written as
\begin{equation}\label{47}
  \left[ \chi_{1} , \chi_{2} \right] = \chi_{12},
\end{equation}
where $T_{12}$ and $Y^{A}_{12}$ are given in Eq.\eqref{44} and
\begin{equation}\label{48}
  \lambda_{12}^{(0)}=\mathcal{L}_{Y_{1}} \lambda_{2}^{(0)}-\mathcal{L}_{Y_{2}} \lambda_{1}^{(0)}.
\end{equation}
Suppose the induced metric on the horizon $\Omega_{AB}$ is conformally related to an off-diagonal one $\gamma_{AB}$, i.e.
\begin{equation}\label{49}
  \Omega_{AB} dx^{A}dx^{B}= \Omega \gamma_{z \bar{z}} dz d \bar{z},
\end{equation}
where $\bar{z}$ is complex conjugate to $z$. For Kerr-Newman (A)dS black holes, $\gamma_{AB}$ describes the Riemann sphere. In this way, the Laurent expansion on the horizon is allowed. The general solution of the conformal Killing equations is $Y= Y^{z}(z) \partial_{z}+Y^{\bar{z}}(\bar{z}) \partial_{\bar{z}}$ and $T= T(z,\bar{z})$ and $\lambda^{(0)}= \lambda^{(0)}(z,\bar{z})$ are arbitrary functions of $z$ and $\bar{z}$. Thus, we can define modes as
\begin{equation}\label{50}
  \begin{split}
       & T_{(m,n)}=\chi (z^{m}\bar{z}^{n},0,0,0), \hspace{1 cm} Y_{m}=\chi (0,-z^{m+1},0,0), \hspace{1 cm} \bar{Y}_{m}=\chi (0,0,-\bar{z}^{m+1},0), \\
       & \lambda^{(0)}_{(m,n)}=\chi (0,0,0,z^{m}\bar{z}^{n}),
  \end{split}
\end{equation}
where $m,n \in \mathbb{Z}$. By using Eq.\eqref{47}, we find the algeba among these modes
\begin{equation}\label{51}
  \begin{split}
       & [Y_{m}, Y_{n}]= (m-n) Y_{m+n}, \hspace{1 cm} [\bar{Y}_{m}, \bar{Y}_{n}]= (m-n) \bar{Y}_{m+n}, \hspace{1 cm} [Y_{m}, \bar{Y}_{n}]=0, \\
       & [T_{(m,n)}, T_{(k,l)}]=0, \hspace{1 cm} [Y_{k}, T_{(m,n)} ]= -m T_{(m+k,n)}, \hspace{1 cm} [\bar{Y}_{k}, T_{(m,n)} ]= -n T_{(m,n+k)},
  \end{split}
\end{equation}
\begin{equation}\label{52}
  \begin{split}
       & [\lambda^{(0)}_{(m,n)}, \lambda^{(0)}_{(k,l)}]=0, \hspace{1 cm} [Y_{k}, \lambda^{(0)}_{(m,n)} ]= -m \lambda^{(0)}_{(m+k,n)}, \hspace{1 cm} [\bar{Y}_{k}, \lambda^{(0)}_{(m,n)} ]= -n \lambda^{(0)}_{(m,n+k)},\\
       & [\lambda^{(0)}_{(m,n)}, T_{(k,l)}]=0,
  \end{split}
\end{equation}
This algebra contains a set of supertranslations current $T_{(m,n)}$ and two sets of Witt algebra currents, given by $Y_{m}$ and $\bar{Y}_{m}$. It also contains a set of multiple-charges current $\lambda^{(0)}_{(m,n)}$. Two sets of Witt currents are in semi-direct sum with the supertranslations and multiple-charges current. The subalgebra \eqref{51} is known as $\mathfrak{bms}_{4}^{H}$ \cite{9} and it differs from Bondi-Metzner-Sachs algebra $\mathfrak{bms}_{4}$ \cite{13,14,15}(the structure constants are different).
\section{Near horizon Charges}\label{S.IV}
Now we find conserved charge conjugate to the asymptotic symmetry generator $\chi$ obtained in previous section. We take codimension-two surface $\mathcal{D}$ in Eq.\eqref{33} to be the horizon
\begin{equation}\label{53}
\begin{split}
   \delta Q_{\chi}= & \oint_{H} \mathcal{Q}_{\text{LW}}^{\mu \nu} [\Phi, \delta \Phi ;\chi ] d^{2}x_{\mu \nu} , \\
    = & \int d^{2} x \mathcal{Q}_{\text{LW}}^{v \rho}\big|_{\rho=0}.
\end{split}
\end{equation}
By substituting the boundary conditions and components of the asymptotic symmetry generators into Eq.\eqref{53}, we have
\begin{equation}\label{54}
  Q_{\chi}= \frac{1}{8 \pi} \int d^{2} x \sqrt{\det \Omega} \left\{ \left(\kappa - \frac{1}{2} \mathcal{N}^{(0)} A_{v}^{(0)} A_{v}^{(1)} \right) T - \frac{1}{2} Y^{A} \left( \theta_{A}+\mathcal{N}^{(0)} A_{v}^{(1)} A_{A}^{(0)} \right) - \frac{1}{2} \mathcal{N}^{(0)} A_{v}^{(1)} \lambda^{(0)}\right\},
\end{equation}
where an integral over one-parameter path on solution phase space was taken. As we mentioned earlier, one can use equations \eqref{21} and \eqref{22} to find the algebra among the conserved charges. After performing some calculations, we find that
\begin{equation}\label{55}
  \left\{ Q_{\chi_{1}}, Q_{\chi_{2}} \right\}= Q_{[\chi_{1}, \chi_{2}]},
\end{equation}
where $[\chi_{1}, \chi_{2}]$ is given by equations \eqref{47}. In this case, by comparing Eq.\eqref{21} and \eqref{55}, we see that the central extension term does not appear. Since the algebra among the conserved charges is isomorphic to the algebra among symmetry generators and the commutation relation \eqref{47} is appeared in the right hand side of Eq.\eqref{55} then it seems reasonable to consider such a commutation relation. By substituting Eq.\eqref{50} into Eq.\eqref{54}, supertranslation, superrotation and multiple-charge modes can be obtained as
\begin{equation}\label{56}
  \mathcal{T}_{(m,n)}= \frac{1}{8 \pi} \int dz d \bar{z} \Omega \sqrt{\gamma} \left(\kappa - \frac{1}{2} \mathcal{N}^{(0)} A_{v}^{(0)} A_{v}^{(1)} \right) z^{m} \bar{z}^{n},
\end{equation}
\begin{equation}\label{57}
    \mathcal{Y}_{m}= \frac{1}{16 \pi} \int dz d \bar{z} \Omega \sqrt{\gamma} \left( \theta_{z}+\mathcal{N}^{(0)} A_{v}^{(1)} A_{z}^{(0)} \right) z^{m+1} ,
\end{equation}
\begin{equation}\label{58}
    \bar{\mathcal{Y}}_{m}=  \frac{1}{16 \pi} \int dz d \bar{z} \Omega \sqrt{\gamma} \left( \theta_{\bar{z}}+\mathcal{N}^{(0)} A_{v}^{(1)} A_{\bar{z}}^{(0)} \right) \bar{z}^{m+1} ,
\end{equation}
\begin{equation}\label{59}
  \mathcal{Q}_{(m,n)}= -\frac{1}{16 \pi} \int dz d \bar{z} \Omega \sqrt{\gamma} \mathcal{N}^{(0)} A_{v}^{(1)} z^{m} \bar{z}^{n} ,
\end{equation}
respectively, where $\gamma = \det (\gamma_{AB})$. Comparing the above equations with Eqs.(6.4), (6.7), (6.25) and (6.26) in Ref.\cite{7}, we deduce that Eq.\eqref{57} and Eq.\eqref{59} will give us the correct value for superrotaion charges and multiple-charges but Eq.\eqref{56} will not give the correct value for supertranslation charges. In fact, we expect that the supertranslation double-zero-mode charge $\mathcal{T}_{(0,0)}$ gives us the black hole entropy multiplied by Hawking temperature $T_{H}= \kappa/2 \pi$.\\
Now, we deviate slightly from the discussion to express the difference between results obtained in this paper and results appeared in Ref.\cite{7}. Compare conserved charge density expression \eqref{31} with one used in Ref.\cite{7} (see Eq.(2.29)) where scalar field was suppressed. They are different! Difference comes from the fact that in obtaining Eq.(2.29) in Ref.\cite{7} the gauge parameter $\lambda$ was redefined as $\lambda+\xi^{\mu} A_{\mu} \rightarrow \lambda$ (see \cite{16} for detailed discussion). This is the origin of the difference.\\
Now let return to our discussion. The equation \eqref{55} gives us the algebra among charge modes
\begin{equation}\label{60}
  \begin{split}
       & \{ \mathcal{Y}_{m}, \mathcal{Y}_{n}\}= (m-n) \mathcal{Y}_{m+n}, \hspace{1 cm} \{ \bar{\mathcal{Y}}_{m}, \bar{\mathcal{Y}}_{n}\}= (m-n) \bar{\mathcal{Y}}_{m+n}, \hspace{1 cm} \{ \mathcal{Y}_{m}, \bar{\mathcal{Y}}_{n}\}=0, \\
       & \{ \mathcal{T}_{(m,n)}, \mathcal{T}_{(k,l)} \}=0, \hspace{1 cm} \{ \mathcal{Y}_{k}, \mathcal{T}_{(m,n)} \} = -m \mathcal{T}_{(m+k,n)}, \hspace{1 cm} \{ \bar{\mathcal{Y}}_{k}, \mathcal{T}_{(m,n)} \}= -n \mathcal{T}_{(m,n+k)},
  \end{split}
\end{equation}
\begin{equation}\label{61}
  \begin{split}
       & \{\mathcal{Q}_{(m,n)}, \mathcal{Q}_{(k,l)}\}=0, \hspace{1 cm} \{\mathcal{Y}_{k}, \mathcal{Q}_{(m,n)} \}= -m \mathcal{Q}_{(m+k,n)}, \hspace{1 cm} \{\bar{\mathcal{Y}}_{k}, \mathcal{Q}_{(m,n)} \}= -n \mathcal{Q}_{(m,n+k)},\\
       & \{\mathcal{Q}_{(m,n)}, \mathcal{T}_{(k,l)}\}=0.
  \end{split}
\end{equation}
In order to find the correct form of supertranslation charges, first we consider algebra among supertranslation modes and multiple-charge modes (see Eq.\eqref{52}). It is clear that $\lambda^{(0)}_{(m,n)}$ and $ T_{(k,l)}$ commute with each other.  By introducing a new mode,
\begin{equation}\label{62}
  \eta_{(m,n)}= (0,0,0,z^{m}\bar{z}^{n}),
\end{equation}
we construct a subalgebra of the algebra \eqref{52}. Therefore, we can define new supertranslation modes as
\begin{equation}\label{63}
  T^{(\text{new})}_{(m,n)} = T_{(m,n)} + \eta_{(m,n)},
\end{equation}
so that they obey same algebra as the old ones do. Strictly speaking, these new modes obey the algebra \eqref{51} and \eqref{52} with $T_{(m,n)} \rightarrow T_{(m,n)}^{(\text{new})}$. Thus, we are allowed to use $U(1)$ gauge fixing to find the correct form of supertranslation charges. To this end, we fix the $U(1)$ gauge freedom as
\begin{equation}\label{64}
  \lambda^{(0)}= -A_{v}^{(0)} T,
\end{equation}
so that corresponding modes are given by Eq.\eqref{62}. In this way, we can define new supertranslation charges conjugate to supertranslation modes  $T_{(m,n)}^{(\text{new})}$ as
\begin{equation}\label{65}
  \mathcal{T}_{(m,n)}^{(\text{new})}= \frac{\kappa}{8 \pi} \int dz d \bar{z} \Omega \sqrt{\gamma} z^{m} \bar{z}^{n}.
\end{equation}
This is exactly what we were looking for. In fact, these are charge modes corresponding to charge conjugate to symmetry generator $\chi = \chi (T, 0, 0, -A_{v}^{(0)} T)$. Eventually, we perform a redefinition as $\tilde{\mathcal{T}}_{(m,n)} = \frac{1}{2 \kappa} \mathcal{T}_{(m,n)}^{(\text{new})}$. In this way, we expect that the supertranslation double-zero-mode $\tilde{\mathcal{T}}_{(0,0)}$ could be related to the black hole entropy as \cite{7,9,10}
\begin{equation}\label{66}
  \mathcal{S}= 4 \pi \tilde{\mathcal{T}}_{(0,0)}.
\end{equation}
Since $\lambda^{(0)}$ is in general a dynamical field independent function and we set it as Eq.\eqref{64} in the last step, i.e. when we want to calculate charges, then $\tilde{\mathcal{T}}_{(0,0)}$ will satisfy the same algebra as \eqref{60} and \eqref{61}. Now, we replace the brackets with commutators, namely $ \{ \hspace{2 mm},\hspace{2 mm} \} \equiv i [\hspace{2 mm},\hspace{2 mm}]$, then the algebra among charge modes becomes
\begin{equation}\label{67}
  \begin{split}
       & i [ \mathcal{Y}_{m}, \mathcal{Y}_{n}]= (m-n) \mathcal{Y}_{m+n}, \hspace{1 cm} i [ \bar{\mathcal{Y}}_{m}, \bar{\mathcal{Y}}_{n}]= (m-n) \bar{\mathcal{Y}}_{m+n}, \\
       & i [ \mathcal{Y}_{k}, \tilde{\mathcal{T}}_{(m,n)} ] = - m \tilde{\mathcal{T}}_{(m+k,n)}, \hspace{1 cm} i [ \bar{\mathcal{Y}}_{k}, \tilde{\mathcal{T}}_{(m,n)} ]= - n \tilde{\mathcal{T}}_{(m,n+k)},\\
       & i [ \mathcal{Y}_{k}, \mathcal{Q}_{(m,n)} ]= - m \mathcal{Q}_{(m+k,n)}, \hspace{1 cm} i [\bar{\mathcal{Y}}_{k}, \mathcal{Q}_{(m,n)} ]= - n \mathcal{Q}_{(m,n+k)}.
  \end{split}
\end{equation}
where commutators not displayed vanish.\\
It is expected that we can apply the Sugawara deconstruction proposed in \cite{17}. To do this, we introduce four new generators $\hat{J}^{\pm}_{m}$ and $\hat{K}^{\pm}_{m}$ so that they obey the following algebra
\begin{equation}\label{68}
  i [\hat{J}^{\pm}_{m}, \hat{K}^{\pm}_{n}]= m \delta_{m+n,0},
\end{equation}
where commutators not displayed vanish. The algebra \eqref{68} consists of two copies of the 3-dimensional flat space near horizon symmetry algebra \cite{18}. Hence we can construct generators $\tilde{\mathcal{T}}_{(m,n)}$ , $\mathcal{Y}^{(\text{new})}_{m}$ and $\bar{\mathcal{Y}}^{(\text{new})}_{m}$ as follows:
\begin{equation}\label{69}
      \tilde{\mathcal{T}}_{(m,n)} = \hat{J}^{+}_{m} \hat{J}^{-}_{n}, \hspace{1 cm}  \mathcal{Y}^{(\text{new})}_{m} = \sum_{p} \hat{J}^{+}_{m-p} \hat{K}^{+}_{p}, \hspace{1 cm} \bar{\mathcal{Y}}^{(\text{new})}_{m} = \sum_{p} \hat{J}^{-}_{m-p} \hat{K}^{-}_{p}.
\end{equation}
It is easy to check that the definitions presented in Eq.\eqref{69} obey the algebra \eqref{67} provided that $\hat{J}^{\pm}_{m}$ and $\hat{K}^{\pm}_{m}$ satisfy the algebra introduced in Eq.\eqref{68}. There will be exist six algebraic constraints on charge zero modes (because we assume that charge zero modes are complex numbers)
\begin{equation}\label{70}
 \frac{\mathcal{S}}{4 \pi} = \hat{J}^{+}_{0} \hat{J}^{-}_{0},\hspace{1 cm}  \pm \frac{i}{2} \mathcal{J} = \hat{J}^{\pm}_{0} \hat{K}^{\pm}_{0},
\end{equation}
where $\mathcal{J}$ is angular momentum of black hole.
\section{Examples}\label{S.V}
Let us consider a typical static dilaton black hole \cite{19,20,21} as first example. The procedure have been done in sections \ref{S.III} and \ref{S.IV} is independent of whether the black object is a black hole or a black string, etc. Therefore, as second expamle, we consider a black string solution of EMD theory. In both examples we set $\Lambda=0$ and $\mathcal{V}=0$.
\subsection{Static Dilaton Black Hole}\label{S.V.A}
Consider a typical static dilaton black hole
\begin{equation}\label{71}
  \begin{split}
     ds^2 = & -F(r) dv^2 +2 dv dr +H(r) (d \theta^{2} + \sin ^{2} \theta d \phi^{2}),  \\
       \varphi = & \varphi_{\infty} + \ln \left| \frac{r+\Sigma}{r-\Sigma}\right| , \hspace{1 cm} A= \left(\frac{e^{\frac{1}{2} \varphi_{\infty}} q}{r-\Sigma} \right) dv,
  \end{split}
\end{equation}
with
\begin{equation}\label{72}
     F(r) = \frac{(r-2M-\Sigma)(r+\Sigma)}{r^{2}-\Sigma^{2}}, \hspace{0.7 cm} H(r)=r^{2}-\Sigma^{2}, \hspace{0.7 cm} \Sigma= -\frac{q^{2}}{2M},
\end{equation}
where $0 \leq r < \infty $, $ 0 \leq \theta \leq \pi$ and $0\leq \phi < 2 \pi$ are radial, polar and azimuthal coordinates, respectvely. This black hole has three parameters $M$, $q$ and $\varphi_{\infty}$ and solves the equations of motion \eqref{4}-\eqref{6} with $\mathcal{N}(\varphi)=4 \exp(-\varphi)$ ($\Lambda=0$ and $\mathcal{V}=0$ were assumed). The event horizon is located at $r_{H}=2M+\Sigma$. Now, we define new radial coordinate $\rho=r-r_{H}$. We can expand metric, $U(1)$ gauge field and scalar field with respect to $\rho$ and find that the corresponding fall-off conditions can be written as Eqs.\eqref{34}-\eqref{36}, where the explicit form of $\kappa$, $A_{v}^{(0)}$ and dynamical fields are given as
\begin{equation}\label{73}
  \begin{split}
       &  \kappa = \frac{1}{4M}, \hspace{0.7 cm} \theta_{A}=0, \hspace{0.7 cm} \Omega_{\theta \theta}=(4M^{2}-2q^{2}),\hspace{0.7 cm} \Omega_{\phi \phi}=(4M^{2}-2q^{2}) \sin ^{2} \theta,   \\
       & \Omega_{\theta \phi}=0,\hspace{0.7 cm} A_{v}^{(0)}=\frac{e^{\frac{1}{2} \varphi_{\infty}} q}{2M}, \hspace{0.7 cm} A_{v}^{(1)}=-\frac{e^{\frac{1}{2} \varphi_{\infty}} q}{4M^{2}},\hspace{0.7 cm} A_{A}^{(0)}=A_{A}^{(1)}=0,\\
       & \varphi^{(0)}=\varphi_{\infty}+ \ln \left| \frac{2M^{2}-q^{2}}{2 M^{2}} \right| .
  \end{split}
\end{equation}
The full 2-metric on horizon,
\begin{equation}\label{74}
  d \sigma ^{2}= (4M^{2}-2q^{2}) \left[ d \theta^{2}+  \sin ^{2} \theta d \phi^{2} \right],
\end{equation}
is conformally related to Riemann sphere. To show this relation, one can introduce a change of coordinates $z= e^{i \phi} \cot \theta/2$ and $\bar{z}= e^{-i \phi} \cot \theta/2$ and show that the conformal factor is given by $\Omega=(4M^{2}-2q^{2})$.\\
Now we can calculate the supertranslation double-zero-mode $\tilde{\mathcal{T}}_{(0,0)}= \frac{1}{4} (4M^{2}-2q^{2})$. Therefore, static dilaton black hole entropy is
\begin{equation}\label{75}
  \mathcal{S}= \pi (4M^{2}-2q^{2}),
\end{equation}
where Eq.\eqref{66} was used. This result is consistent with the previous result obtained in \cite{22}. It is clear that superrotation charge modes $\mathcal{Y}_{m}$ and $\bar{\mathcal{Y}}_{m}$ are zero. By substituting Eq.\eqref{73} and Eq.\eqref{74} into Eq.\eqref{59}, one can find multiple-charge double-zero-mode as
\begin{equation}\label{76}
  \mathcal{Q}_{(0,0)}=q e^{-\frac{1}{2} \varphi_{\infty}},
\end{equation}
which is exactly total electric charge of static dilaton black hole \cite{22}.\\
The given black hole has an inner and outer horizons of radius $r_{-}= -\Sigma$ and $r_{+}=2M+\Sigma$, respectively. One can choose zero mode eigenvalues of $\hat{J}^{\pm}_{m}$ and $\hat{K}^{\pm}_{m}$ as
\begin{equation}\label{77}
  \hat{J}^{\pm}_{0}=\frac{1}{2}(r_{+}\pm i r_{-}), \hspace{0.7 cm} \hat{K}^{\pm}_{0}=0,
\end{equation}
so that they satisfy Eq.\eqref{70}.
\subsection{Rotating Charged Dilaton Black String}\label{S.V.B}
Now consider a rotating charged dilaton black string \cite{27}
\begin{equation}\label{78}
  ds^2 =  -F(r) \left( \Xi dt - a d \hat{\phi} \right)^{2} + \frac{dr^{2}}{F(r)} + R(r) \left( \frac{a}{l^{2}} dt - \Xi d \hat{\phi} \right)^{2} + R(r) \frac{dx^{2}}{l^{2}},
\end{equation}
\begin{equation}\label{79}
       \varphi = \frac{\beta}{\alpha} \ln \left( \frac{b}{r} \right) , \hspace{1 cm} A=- \frac{q}{r} \left(\Xi dt - a d \hat{\phi} \right),
\end{equation}
with
\begin{equation}\label{80}
     F(r) = r^{\beta - 1} \left( -M + \frac{(1+\alpha^{2}) q^{2}}{b^{\beta} r}\right), \hspace{0.7 cm} R(r)=b^{\beta} r^{2-\beta}, \hspace{0.7 cm} \Xi^{2}= 1+\frac{a^{2}}{l^{2}},\hspace{0.7 cm} \beta= \frac{2 \alpha^{2}}{1+\alpha^{2}},
\end{equation}
where $- \infty < t < \infty$ is time coordinate. The line-element \eqref{78} describes a black string for $0\leq \hat{\phi} < 2 \pi$ and $- \infty <x< \infty$. This black string has three parameters $M > 0$, $q$ and $b$ and solves the equations of motion \eqref{4}-\eqref{6} with $\mathcal{N}(\varphi)=4 \exp(- \alpha \varphi)$, where $\alpha$ is a constant. Also, $l$ is a constant and it has dimension of length. The spacetime described by Eq.\eqref{78}  presents a naked singularity with a regular cosmological horizon at
\begin{equation}\label{81}
  r_{H}= \frac{(1+\alpha^{2}) q^{2}}{b^{\beta} M}.
\end{equation}
In order to find near horizon geometry of rotating charged dilaton black string, first we write the metric \eqref{78} in the advanced Eddington-Finkelstein coordinates $(v,r, \tilde{\phi},x)$. To this end, we transform coordinates as
\begin{equation}\label{82}
  dv=dt+ \frac{1- B(r) g_{t \hat{\phi}}}{g_{tt}} dr , \hspace{1 cm} d \tilde{\phi}= d \hat{\phi}+ B(r) dr,
\end{equation}
with
\begin{equation}\label{83}
  B(r)=\frac{a}{F(r) R(r) l^{2}} \sqrt{R(r) [R(r)-l^{2}F(r)]}.
\end{equation}
Next, we perform another coordinates transformation as $\tilde{\phi}= \phi + \Omega_{H} v$, where $\Omega_{H}= \frac{a}{\Xi l^{2}}$ is the horizon velocity of rotating charged dilaton black string. Finally, we find that
\begin{equation}\label{84}
\begin{split}
   ds^{2}= & -\frac{F}{\Xi^{2}} dv^{2} + \frac{2 F \left( 1+\frac{a}{\Xi l^{2}} R B\right)}{ \left(F \Xi^{2}-\frac{a^{2}}{l^{4}}R \right)} dv dr+ \frac{2 a F}{\Xi} dv d \phi +\frac{2 F R B \left( 1+\frac{\Xi l^{2}}{a} F B\right)}{ \left(F \Xi^{2}-\frac{a^{2}}{l^{4}}R \right)} dr d\phi \\
     & + \left( - F a^{2}+ R\Xi^{2} \right) d \phi^{2}+ R\frac{dx^{2}}{l^{2}}.
\end{split}
\end{equation}
In this coordinate system the $U(1)$ gauge field can be written as
\begin{equation}\label{85}
  A=-\frac{q}{r} \left(\frac{dv}{\Xi} -a d \phi \right),
\end{equation}
and scalar field remains unchanged. Also, we have $g_{vv} =g_{v \phi} =0$ on the cosmological horizon. Now we write near horizon fall-off conditions for rotating charged dilaton black string in the Gaussian null coordinate system. To do this, we follow the method proposed in Appendix A of the paper \cite{24}. Therefore, we rewrite the metric relative to the correct set of geodesics. A suitable pair of cross-normalized null normals is
\begin{equation}\label{86}
  l= \partial_{v}, \hspace{1 cm} \text{and} \hspace{1 cm} n= \left(\frac{l^{4} (1+\Xi)^{2}}{2 a^{2} b^{\beta} r_{H}^{2-\beta}} \right) \partial _{v}- \Xi \partial _{r}- \left(\frac{l^{2} (1+\Xi)}{ a b^{\beta} r_{H}^{2-\beta} \Xi}\right) \partial_{\phi}.
\end{equation}
These vectors are defined on horizon and we have $l \cdot l |_{H}=n \cdot n |_{H}=0$ and $l \cdot n =1$. Now we consider a family of null geodesics that crosses $H$. The vector field tangent to them is $n$ and they are labeled by $(v, \theta, \phi)$. Suppose $\rho$ is an affine parameter which parameterize the given geodesics so that $\rho=0$ on $H$. The geodesics can be constructed up to second order in $\rho$:
\begin{equation}\label{87}
  X_{(v,\theta, \phi)}^{\mu}(\rho)= X^{\mu}\big|_{\rho =0} + \rho \frac{d X^{\mu}}{d \rho} \bigg|_{\rho=0}+ \mathcal{O}(\rho^{2}),
\end{equation}
where $X^{\mu}\big|_{\rho =0} = (v,r_{H},\theta, \phi)$ and $\frac{d X^{\mu}}{d \rho} \big|_{\rho=0}=n^{\mu}$. The equation \eqref{87} defines a transformation from $(v,r,\theta, \phi)$ to $(v, \rho, \theta , \phi)$ and then we can calculate the first order expansion of the metric $g_{\mu \nu} = g_{\mu \nu}^{(0)}+ \rho g_{\mu \nu}^{(1)}+ \mathcal{O}(\rho^{2})$, where
\begin{equation}\label{88}
  g_{v \rho}^{(0)}= 1, \hspace{1 cm} g_{\phi \phi}^{(0)}= b^{\beta} r_{H}^{2-\beta} \Xi^{2} , \hspace{1 cm} g_{x x}^{(0)}= \frac{b^{\beta} r_{H}^{2-\beta}}{l^{2}},
\end{equation}
\begin{equation}\label{89}
  g_{vv}^{(1)}=-2 \kappa , \hspace{0.8 cm} g_{v \phi}^{(1)}= a M r_{H}^{\beta-2},
\end{equation}
where $\kappa$ is surface gravity of the rotating charged dilaton black string,
\begin{equation}\label{90}
  \kappa = \frac{M r_{H}^{\beta-2}}{2 \Xi}.
\end{equation}
In the new coordinate system, the $U(1)$gauge field and scalar field can be expanded as
\begin{equation}\label{91}
\begin{split}
     & A_{v}=- \frac{q}{r_{H} \Xi} - \frac{q}{r_{H}^{2}} \rho +\mathcal{O}(\rho^{2}), \hspace{1 cm} A_{\phi}=\frac{a q}{r_{H}} + \frac{a q \Xi}{r_{H}^{2}}+\mathcal{O}(\rho^{2}),\\
     &  A_{\rho}=\mathcal{O}(\rho^{2}), \hspace{1 cm} A_{x}= \mathcal{O}(\rho^{2}) , \\
     & \varphi = \frac{\beta}{\alpha} \ln \left( \frac{b}{r_{H}} \right)+ \frac{\beta \Xi}{\alpha r_{H}} \rho +\mathcal{O}(\rho^{2}).
\end{split}
\end{equation}
From Eq.\eqref{88}, the full 2-metric on horizon is
\begin{equation}\label{92}
  d \sigma ^{2}= b^{\beta} r_{H}^{2-\beta} \left( \Xi^{2} d \phi^{2} + \frac{dx^{2}}{l^{2}} \right).
\end{equation}
This metric is conformally related to cylinder. To show this relation, we introduce a field-dependent change of coordinates
\begin{equation}\label{93}
  z= \exp \left( \frac{x}{\Xi l}+ i \phi\right), \hspace{1 cm} \bar{z}= \exp \left( \frac{x}{\Xi l}- i \phi\right).
\end{equation}
Now, we can write the metric of the horizon in the conformal form
\begin{equation}\label{94}
\begin{split}
    d \sigma^{2} =& \Omega \gamma_{AB} dx^{A} dx^{B} \\
     =& \Omega dz d\bar{z},
\end{split}
\end{equation}
with
\begin{equation}\label{95}
  \Omega= b^{\beta} r_{H}^{2-\beta} \Xi^{2} \exp \left(-\frac{2x}{\Xi l} \right).
\end{equation}
Here, $\gamma_{AB}$ is metric on a cylinder. The conformal factor $\Omega$ is a function of $z$ and $\bar{z}$. The induced metric on the horizon is locally, conformally equivalent to a cylinder and hence the Laurent expansion on the cylinder is allowed. Therefore, we can use Eqs.\eqref{65} and \eqref{57}-\eqref{59} to find charge modes on the horizon of a rotating charged black string.\\
The supertranslation double-zero-mode charge per unit length of the string can be obtained as
\begin{equation}\label{96}
  \tilde{\mathcal{T}}_{(0,0)}=\frac{b^{\beta} r_{H}^{2-\beta} \Xi}{8 l},
\end{equation}
and hence the entropy per unit length of the string will be
\begin{equation}\label{97}
  \mathcal{S}=\frac{\pi b^{\beta} r_{H}^{2-\beta} \Xi}{2 l},
\end{equation}
where Eq.\eqref{66} was used. Also, one can show that superrotation charge and multiple-charge modes per unit length of the string are
\begin{equation}\label{98}
  \begin{split}
      & \mathcal{Y}_{m} = \frac{i a M \Xi b^{\beta}}{16 l} \left( \frac{3- \alpha^{2}}{1+\alpha^{2}}\right) \delta_{m,0}, \hspace{0.7 cm} \bar{\mathcal{Y}}_{m} =- \frac{i a M \Xi b^{\beta}}{16 l} \left( \frac{3- \alpha^{2}}{1+\alpha^{2}}\right) \delta_{m,0}, \\
       & \mathcal{Q}_{(0,0)}= \frac{q \Xi}{2l},
  \end{split}
\end{equation}
respectively. Therefore, angular momentum and electric charge per unit length of the string can be read as
\begin{equation}\label{99}
  \mathcal{J}= \frac{ a M \Xi b^{\beta}}{8 l} \left( \frac{3- \alpha^{2}}{1+\alpha^{2}}\right), \hspace{0.7 cm} \mathcal{Q}_{E}=\frac{q \Xi}{2l},
\end{equation}
respectively. One can choose zero mode eigenvalues of $\hat{J}^{\pm}_{m}$ and $\hat{K}^{\pm}_{m}$ as
\begin{equation}\label{100}
  \hat{J}^{\pm}_{0}= \left(\frac{b^{\beta} r_{H}^{2-\beta}}{8 l \Xi} \right)^{\frac{1}{2}} (1 \pm \frac{ia}{l}), \hspace{0.7 cm} \hat{K}^{\pm}_{0}=-\left(\frac{b^{\beta} r_{H}^{2-\beta}}{8 l \Xi} \right)^{-\frac{1}{2}} \left( \frac{3- \alpha^{2}}{1+\alpha^{2}}\right) \left( \frac{a^{2} b^{\beta} M}{16l^{2} \Xi}\right) (1 \pm \frac{il}{a}),
\end{equation}
so that they satisfy Eq.\eqref{70}. The results appeared in Eq.\eqref{97} and Eq.\eqref{99} are coincident to the results that was obtained in Ref.\cite{27}.
\section{Conclusion}
In this paper, we have considered Einstein-Maxwell-Dilaton (EMD) theory. In EMD theory, by using Eq.\eqref{31}, conserved charge conjugate to symmetry generator $\chi= (\xi, \lambda)$ can be obtained. The expression \eqref{31} differs from the one obtained in Ref.\cite{16} and used in \cite{7}. Difference comes from the fact that in obtaining Eq.(2.29) in Refs.\cite{7,16}, the gauge parameter $\lambda$ was redefined as $\lambda+\xi^{\mu} A_{\mu} \rightarrow \lambda$. In section \ref{S.III}, we have considered near horizon fall-off conditions for metric, $U(1)$ gauge field and scalar field in the Gaussian null coordinate system. Equations of motion implied that surface gravity $\kappa$ and first order term of timelike component of $U(1)$ gauge field $A^{(0)}_{v}$ have to be constants for stationary black holes. components of symmetry generator $\chi$ are given by Eq.\eqref{37} and Eq.\eqref{38}. The change in dynamical fields are given by Eqs.\eqref{39}-\eqref{40} under the action of symmetry generator $\chi$.  We have assumed that the induced metric on the horizon $\Omega_{AB}$ is conformally related to an off-diagonal one $\gamma_{AB}$ (where $x^{A}=\{ z , \bar{z}\}$ and $\bar{z}$ is complex conjugate to $z$) and hence the Laurent expansion on the horizon is allowed. Supertranslation, superrotation and multiple-charge modes are given by Eq.\eqref{50} and the algebra among them contains a set of supertranslations current $T_{(m,n)}$ and two sets of Witt algebra currents, given by $Y_{m}$ and $\bar{Y}_{m}$. It also contains a set of multiple-charges current $\lambda^{(0)}_{(m,n)}$. Two sets of Witt currents are in semi-direct sum with the supertranslations and multiple-charges current. We have obtained near horizon conserved charge conjugate to near horizon symmetry generator $\chi$ and have shown that the algebra among the conserved charges is given by Eq.\eqref{55}. Consequently, equations \eqref{56}-\eqref{59}, give us charge modes. we expect that the supertranslation double-zero-mode charge $\mathcal{T}_{(0,0)}$ gives us the black hole entropy multiplied by Hawking temperature. Because of the presence of second term in integrand in Eq.\eqref{56} this statement could not be true. This may be seen as a problem but it is not. Because, in EMD theory, we have a $U(1)$ gauge freedom and we can use of an appropriate gauge fixing to fix that problem. To this end, we have introduced new supertranslation modes $T^{\text{(new)}}_{(m,n)}$, see Eq.\eqref{63}, and fixed the $U(1)$ gauge freedom as Eq.\eqref{64}. Supertranslation charges \eqref{65} conjugate to these new supertranslation modes are exactly what we were looking for. By redefining new Supertranslation charges as $\tilde{\mathcal{T}}_{(m,n)} = \frac{1}{2 \kappa} \mathcal{T}_{(m,n)}^{(\text{new})}$, the relation between black hole entropy $\mathcal{S}$ and $\tilde{\mathcal{T}}_{(0,0)}$ is given by Eq.\eqref{66}. We have shown that $\tilde{\mathcal{T}}_{(m,n)}$ together with superrotation charge modes and multiple-charge modes satisfy the algebra \eqref{67}. By introducing four new generators $\hat{J}^{\pm}_{m}$ and $\hat{K}^{\pm}_{m}$ so that they obey the algebra \eqref{68}, one can apply the Sugawara deconstruction proposed in \cite{17}. In this way, one can construct generators $\tilde{\mathcal{T}}_{(m,n)}$ , $\mathcal{Y}^{(\text{new})}_{m}$ and $\bar{\mathcal{Y}}^{(\text{new})}_{m}$ as \eqref{69}. Hence, the relation among zero modes are given by Eq.\eqref{70}, where $\mathcal{J}$ is angular momentum of black hole. In section \ref{S.V}, we have considered two examples, one a typical static dilaton black hole and another a charged rotating black string. Because the procedure have been done in sections \ref{S.III} and \ref{S.IV} is independent of whether the black object is a black hole or a black string then, we can apply the method on a black string as well as a black hole. In both examples we have assume that $\Lambda= \mathcal{V}=0$. For a typical static dilaton black hole, the full 2-metric on horizon is conformally related to Riemann sphere. Using \eqref{56}-\eqref{59} we have found charge zero modes. The obtained results are exactly matched with the results of previous works. In subsection \ref{S.V.B}, we wrote near horizon fall-off-conditions correspond to rotating charged dilaton black string in the Gaussian null coordinate system (see equations \eqref{88}-\eqref{91}). We have shown that the induced metric on the horizon of rotating charged dilaton black string is locally, conformally equivalent to cylinder. Therefore, we have used Eqs.\eqref{65} and \eqref{57}-\eqref{59} to find the value of charge zero modes per unit length of the string. Using these two examples, we showed that the new entropy formula proposed in \cite{17} is valid in EMD theory and for black strings.

\section{Acknowledgments}
The work of Hamed Adami has been financially supported by Research Institute for Astronomy Astrophysics of Maragha (RIAAM).

\end{document}